\documentclass[a4paper]{article}

\usepackage{ICASSP2021}
\usepackage{comment}
\usepackage{adjustbox}
\usepackage{amsmath}
\title{Unsupervised Spoken Term Discovery Based on Re-clustering of Hypothesized Speech Segments with Siamese and Triplet Networks}

\name{Man-Ling Sung, Tan Lee}
\address{Department of Electronic Engineering,
The Chinese University of Hong Kong}
\email{mlsung@link.cuhk.edu.hk, tanlee@ee.cuhk.edu.hk}

\begin{document}
\ninept
\maketitle
\begin{abstract}
Spoken term discovery from untranscribed speech audio could be achieved via a two-stage process. In the first stage, the unlabelled speech is decoded into a sequence of subword units that are learned and modelled in an unsupervised manner. In the second stage, partial sequence matching and clustering are performed on the decoded subword sequences, resulting in a set of discovered words or phrases. A limitation of this approach is that the results of subword decoding could be erroneous, and the errors would impact the subsequent steps. While Siamese/Triplet network is one approach to learn segment representations that can improve the discovery process, the challenge in spoken term discovery under a complete unsupervised scenario is that training examples are unavailable. In this paper, we propose to generate training examples from initial hypothesized sequence clusters. The Siamese/Triplet network is trained on the hypothesized examples to measure the similarity between two speech segments and hereby perform re-clustering of all hypothesized subword sequences to achieve spoken term discovery. 
Experimental results show that the proposed approach is effective in obtaining training examples for Siamese and Triplet networks, improving the efficacy of spoken term discovery as compared with the original two-stage method.

\end{abstract}

\noindent\textbf{Index Terms}: 
Spoken term discovery, Siamese network, Triplet network, segment representation learning, zero resource speech technology

\section{Introduction}
%What is spoken term detection 
Unsupervised speech modeling is the task of discovering and modeling speech units at various levels from audio recording without using any prior linguistic information.
It is an interesting, challenging and impactful research problem as phonetic, lexical and even semantic information could be acquired without the process of transcribing and understanding the given speech data. The relevant technology is particularly important to facilitate data preparation especially in the scenarios where: 1) a large (even unlimited) amount of audio data are readily available online but they are untranscribed; 2) a large amount of audio recording is available for a smaller language which no structured linguistic knowledge or documentation can be found.

Spoken term discovery is a representative task of unsupervised speech modeling. It aims to discover repetitively occurred words and/or phrases from untranscribed audio. 
The problem is commonly tackled with a two-stage approach. In the first stage, a set of subword units are automatically discovered from untranscribed speech data and these units in turn can be used to represent the speech data as a symbol sequence. In the second stage, variable-length sequence matching and clustering are performed on the subword sequence representations. One major drawback of this is that the subword decoding errors in the first stage would propagate to deteriorate the outcome of spoken term discovery in the second stage.
The present study investigates the use of Siamese and Triplet networks in learning segment representations for spoken term discovery when no training labels are available. Siamese network has been commonly applied to pattern classification or matching problems when only weak labels are available. We propose to train a Siamese/Triplet network with a small dataset of matched and mismatched sequence pairs obtained and use the trained network to generate feature representations for unseen subword sequences. The training dataset is constructed based on hypothesized spoken term clusters from an baseline spoken term discovery system developed in our previous study. With the new feature representations learned by the Siamese/Triplet network, re-clustering of subword sequences is carried out to generate an improved set of discovered spoken terms.

\section{Related Work}

\subsection{Spoken term discovery}

The first attempt in discovering acoustic units is the acoustic segment model (ASM) \cite{wilpon1987investigation, lee1988segment}, where a self-derived acoustic model is trained on untranscribed audio to discover subword units. There are attempts that further extended to discover longer acoustic units such as spoken term segments in \cite{park2005towards} through pattern matching of acoustic features, which is later known as spoken term discovery. Spoken term discovery aims to find and extract repetitively occurred sequential pattern from audio in an unsupervised manner.
%There are different ways of implementation. 
In general, a spoken term discovery system performs three tasks one after the other: segmentation, matching and clustering \cite{versteegh2015zero}. 

%The repeated patterns are then clustered to form the spoken terms. 

There are mainly two approaches to spoken term discovery. In the first approach, pattern discovery is done directly with acoustic features. Word-level speech segments are matched using sequence matching algorithms like segmental-DTW. The matching could be based on conventional frame-level features \cite{park2005towards} or fixed-dimension segment representations \cite{kamper2017embedded, thual2018k}. Another approach involves a two-stage process. The ASM is trained on untranscribed audio, resulting in symbolic representations known as the pseudo-transcription of speech. 
Sequential pattern discovery is then performed by local alignment or string matching and clustering of sequential patterns \cite{jansen2011efficient, harwath2013zero, siu2014unsupervised, sung2018unsupervised}. The results of clustering could be corresponded to the discovered spoken terms in the given audio dataset.

\subsection{Siamese and Triplet networks}

Siamese neural network was proposed in \cite{bromley1994signature}. It consists of two identical sub-network components, which share the learnable parameters. Through the two sub-network components, Siamese neural network is trained to perform a designated classification task on a pair of data samples. The most common task is to determine whether the two input samples are from the same class or not. In other words, the exact class identities for individual training samples are not needed. The training of Siamese network requires relatively fewer training samples than conventional neural network classifiers \cite{koch2015siamese}. Siamese network is widely used in computer vision. It is shown to have the ability of comparing samples from unseen classes in the problem of one shot classification \cite{koch2015siamese}. 
Triplet network \cite{hoffer2015deep} is an extension of Siamese network. It consists of three identical sub-networks, which process 3 input samples in parallel, including one reference sample, one matched and one mismatched samples. The network is trained to capture the similarity between the matched sample and the reference and the dissimilarity between the mismatched sample and the reference.

\subsection{Siamese network on spoken term detection/discovery}

%Siamese network has been used in learning speech embeddings and it has shown to be able to learn effective subword units and term units representations (cite).
It has been shown that Siamese network is able to learn new representations from audio signals, which facilitate spoken word classification \cite{kamper2016deep}. 
It is also able to generate effective representations for spoken term detection \cite{svec2017relevance, zhu2018siamese}. While existing work assumes matched pair and mismatched pairs for training the Siamese network are available, one challenge in unsupervised spoken term discovery is that no information is given to the system except the audio only. In order to apply Siamese network in learning segment representations, reliable matched and mismatched pairs are required for training the network. 

Relatively less work is done on unsupervised generation of matched and mismatched training pairs. There is work that identifies frame-level training samples. After segmentation, frames from same segments are treated as matched pairs, frames from adjacent segments are treated as mismatched pairs \cite{bhati2019unsupervised}. There is also work that extracts training examples from available spoken term discovery system, with sampling based on distributions of speakers and matched/mismatched pairs \cite{riad2018sampling}.

%“Unsupervised Acoustic Segmentation and Clustering using Siamese Network Embeddings” 
%Pairs of frames from the same segment are considered as matched pairs, while pairs of frames from adjacent segments are considered as mismatched pairs.
%frame level
%---> match: from same segments, mismatch: adjacent segments

%“Sampling strategies in Siamese Networks for unsupervised speech representation learning“
% We use the orthographic transcription from word-level annotations to determine same and different pairs to train the siamese networks. In the fully unsupervised setting, we obtain pairs of same and different words from the Track 2 baseline of the 2015 ZeroSpeech challenge [3]: the Spoken Term Discovery system from [13]. We use both the original files from the baseline, and a rerun of the algorithm with systematic variations on its similarity threshold parameter
% they only use one line to mention the similarity filter --> maybe can eleborate on it

\section{Proposed System}

To generate reliable matched and mismatched pair, we consider the approach of relaying information generated from a trained spoken term discovery system. Subwords and term clusters are learned in unsupervised manner, training pairs are identified by evaluating the discovered term clusters based on the discovered subword units.

In our previous work \cite{sung2018unsupervised}, a two-stage spoken term discovery approach was investigated on recording of classroom lectures. The audio signals are first converted into frame-level bottleneck features using a multilingual deep neural network model. A set of subword-level speech units are discovered based on the bottleneck features. The discovered subword units are treated as phonemes to be the acoustic modeling units in a conventional ASR system. The audio signals are in turn decoded by the ASR system into pseudo-transcription. Sequential pattern matching is applied to the pseudo-transcription to obtain segments represented in subword sequences, follow by clustering of the subword sequences. The resulted clusters were shown to be strongly associated with keywords or key phrases that occur frequently in the audio signals. In particular, clusters formed by long subword sequences generally are able to represent meaningful whole words or phrases. Nevertheless, many of the resulted clusters, especially those formed by short sequences, do not provide much useful information for spoken term discovery.

\subsection{Training data for Siamese network}
\label{sec:conf}
The intended problem of spoken term discovery assumes the absence of any kind of data labels for supervised model training. To address this issue, we adopt the Siamese/Triplet network, which can be trained with weakly labelled data to learn robust segment-level representation of speech. The required segments and their ``weak'' labels, which tell whether a pair of speech segments contain the same or different spoken terms, is obtained by leveraging the preliminary clustering results of the two-stage approach described above. Simply speaking, the clusters with high ``purity'' are used to provide the training data and their labels.

Let $C$ denote a cluster initially determined by the two-stage approach. $C$ contains a number of speech segments that hypothetically correspond to the same word or phrase. Consider two segments $i$ and $j$ in $C$, and let $lev(i, j)$ be the Levenshtein distance between the symbol representations of $i$ and $j$. We calculate the mean and standard deviation of the Levenshtein distances of all pairs of segments in $C$, i.e.,

\begin{equation}
\mu_s = \sum_{i,j \in C} {lev(i, j) \over |C|^2}
\end{equation}
\begin{equation}
\sigma_s = \sqrt{{\sum_{i,j \in C} (lev(i, j) - \mu_s)^2 \over |C|^2}}.
\end{equation}
A small value of $\mu_s$ implies that members in $C$ have similar pseudo representations. A small value of $\sigma_s$ means that the distances between different member pairs are similar. These two measures can be used to indicate the purity of $C$. We propose to retain a set of clusters with $\mu_s$ and $\sigma_s$ below certain empirically determined thresholds, i.e.,
\begin{align*}
C \in T \text{ if }
\mu_s < thres_{\mu_s} \cdot \bar{C}
\text{ and }
\sigma_s < thres_{\sigma_s} \cdot \bar{C},
\end{align*}
where $\bar{C}$ denotes the average length of symbol sequences in $C$.
%For all initial clusters hypothesized by the two-stage approach, the intra-cluster distances are computed. Those with small intra-cluster distances are retained and denote as the set $T$. To obtain $T$, we measure the confidence of the groups that are mostly likely to be clustered correctly.

%$\mu < 0.2 \times |C| $ and $\sigma < 0.15 \times |C|$
%Among the clusters in $T$, samples in the same clusters are used for match pairs,
%To obtain the similar/dissimilar spoken term pairs from $T$ for the training set.
%calculate large set of Levenshtein distance.
%automata \cite{schulz2002fast}.
%\cite{jhaver2014calculating}

Let the collection of retained ``pure'' clusters be denoted by $T$. Speech segments in the same cluster are believed to contain the same spoken term and therefore are used to form matching pairs for the training of Siamese/Triplet network. On the other hand, contrasting training pairs are formed by segments from contrasting clusters that have large inter-cluster distance. Consider clusters $C_1$ and $C_2$ in $T$, and define
\begin{equation}
\mu_d = \sum_{i \in C_1, j \in C_2} {lev(i, j) \over |C_1|  |C_2|}
\end{equation}
\begin{equation}
\sigma_d = \sqrt{{\sum_{i \in C_1, j \in C_2} (lev(i, j) - \mu_d)^2 \over |C_1|  |C_2|}}.
\end{equation}
$C_1$ and $C_2$ are selected as contrasting clusters if
\begin{align*}
    \mu_d >  thres_{\mu_d} \cdot {(\bar{C_1} + \bar{C_2}) \over 2} 
    \text{ and } 
    \sigma_d < thres_{\sigma_d} \cdot {(\bar{C_1} + \bar{C_2}) \over 2}
\end{align*}

% \begin{align*}
%     \mu_d >  thres_{\mu_d} \times & {(|C_1| + |C_2|) \over 2} \\
%     \text{ and } & \\
%     \sigma_d < thres_{\sigma_d} \times & {(|C_1| + |C_2|) \over 2}
% \end{align*}

\subsection{Siamese/Triplet network}

%(describe Siamese network settings, )

%\subsection{Models}

The Siamese network consists of two identical convolution neural networks (CNN) with shared parameters. 
In the proposed model, the two CNN take in the bottleneck features from a pair of speech segments, denoted $x_0$ and $x_1$, and their outputs are the respective learnt representations denoted as $f(x_0)$ and $f(x_1)$. %Different from traditional classification network, only soft labelling is needed. 
If $x_0$ and $x_1$ are a matched pair, the overall output of the Siamese network is trained to be $1$. If they are a mismatched pair, the output is trained to be $0$.

\begin{figure}[ht]
    \centering
    \includegraphics[scale=0.35]{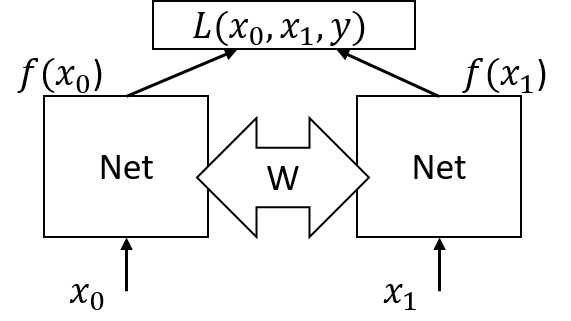}%\vspace{-0.1cm}
    \includegraphics[scale=0.35]{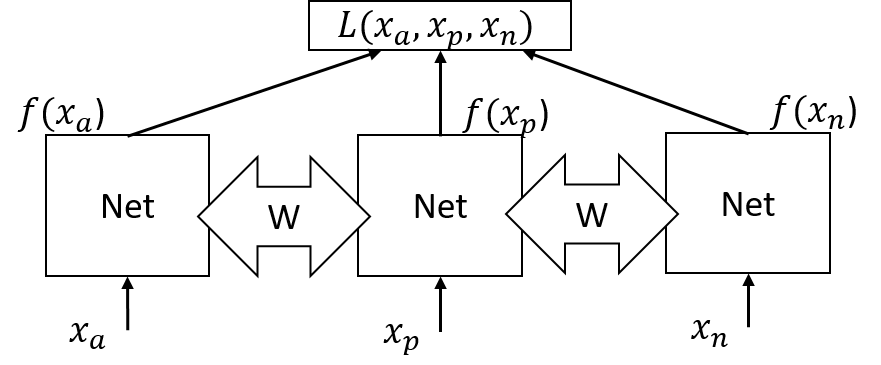}
    \caption{Siamese network (left) and Triplet network (right)}
    \label{fig:siamese}
\end{figure} 
%\vspace{-0.1cm}
% \begin{figure}[ht]
%     \centering
%     \includegraphics[scale=0.35]{image/siamese_network.png}%\vspace{-0.1cm}
%     \caption{Siamese network}
%     \label{fig:siamese}
% \end{figure} 
% % %\vspace{-0.35cm}

The network parameters are trained to minimize the contrastive loss function defined as  %$L(x_0, x_1,y)  =$
% \begin{equation}
% %\begin{aligned}
%  L(x_0, x_1,y)  = &  {1 \over 2} \{ y || f(x_0) - f(x_1)||^2_2 +  \\
% &    (1-y) max(0, m - ||f(x_0) - f(x_1)||^2_2  \}\\
% %\end{aligned}
% \end{equation}
% ${1 \over 2} \{ y || f(x_0) - f(x_1)||^2_2 +  
 % (1-y) max(0, m - ||f(x_0) - f(x_1)||^2_2  \}$, where $m$ is the margin.
\begin{equation}
\begin{aligned}
 L(x_0,& x_1,y) = {1 \over 2} y || f(x_0) - f(x_1)||^2_2 +  \\
&     {1 \over 2} (1-y) \{max(0, m - ||f(x_0) - f(x_1)||_2\}^2
\end{aligned}
\end{equation}

The Triplet network is also composed of the same type of CNN components. It takes three segments $x_a$, $x_p$ and $x_n$ as the input, where $x_p$ and $x_a$ are matched pair, and $x_n$ and $x_a$ are mismatched pair. %the learnt output of same and different pairs $\{x_a, x_p\}$ and $\{x_a, x_n\}$ are compared at the same time. 
%The weights of the three network components are shared.
% %\vspace{-0.1cm}
% \begin{figure}[ht]
%     \centering
%     \includegraphics[scale=0.35]{image/triplet_network.png} %\vspace{-0.1cm}
%     \caption{Triplet network}
%     \label{fig:triplet}
% \end{figure}
% %\vspace{-0.35cm}
The Triplet network aims at embedding matching samples closer and meanwhile keeping contrasting samples away in the representation space. The Triplet loss function is given as,
\begin{equation}
\begin{aligned}
L &(x_a,x_p,x_n) =\\  
& max(0, m+ ||f(x_a) - f(x_p)||^2_2 - 
 || f(x_a) - f(x_n)||^2_2)
\end{aligned}
\end{equation}
%$L(x_a,x_p,x_n) =
% max(0, m+ ||f(x_a) - f(x_p)||^2_2 -  || f(x_a) - f(x_n)||^2_2)$,
where $m$ denotes the the margin between matched and mismatched samples from $x_a$.

\subsection{Segment representations clustering}
%https://towardsdatascience.com/how-dbscan-works-and-why-should-i-use-it-443b4a191c80

The Siamese/Triplet network is trained to learn segment representations that can be used to measure the similarity between segments. Our idea is to apply re-clustering on speech segments so as to achieve spoken term discovery. Hierarchical Density-based Spatial clustering of Applications with Noise (HDBSCAN) \cite{campello2013density} is adopted. The clustering algorithm uses data samples to construct a minimum spanning tree of the distance-weighted graph. Each node of the tree represents a data sample, and the weight of edge connecting two nodes represents the distance between the data samples. A hierarchical level of clustering is built from the tree. The tree is then condensed based on minimal cluster size and finally stable clusters are obtained.

In some cases, HDBSCAN may produce lot of micro-clusters on high-density region, so methods that combine the use of DBSCAN and HDBSCAN are introduced, such as introducing cluster selection epsilon that extracts DBSCAN results on region larger than the epsilon instead \cite{malzer2019hybrid}. This hybrid clustering approach is also considered in our implementation.

\section{Experimental Setup}

\begin{table*}[t]
\resizebox{17cm}{!}{%
\begin{tabular}{lcccccccccccccccc}
\hline
 \textbf{English}& \multicolumn{3}{c}{\textbf{Grouping}}& \multicolumn{3}{c}{\textbf{Token}}& \multicolumn{3}{c}{\textbf{Type}} & \multicolumn{3}{c}{\textbf{Boundary}}& \multicolumn{4}{c}{\textbf{NLP}} \\ \hline
   & P  & R  & F & P & R & F & P & R & F & P  & R  & F & NED  & Cov & n-words & n-pairs \\ \hline
Baseline  & NA& NA & NA & 3.5  & 5.9   & 4.4  & \textbf{1.8} & \textbf{22.1} & \textbf{3.3} & 26.6  & 44.7  & 33.4  & 92.9  & \textbf{78.2} & 260275  & 126902906 \\
Siamese (hybrid)  & NA& NA & NA & 3.4  & 8.0   & 4.8  & 1.8  & 19.5  & 3.3  & 26.4  & 43.5  & 32.9  & 94.6  & 73.1  & 226466  & 18046322  \\
Triplet (hybrid)  & NA& NA & NA & \textbf{3.3} & \textbf{10.1} & \textbf{5.0} & 1.8  & 20.0  & 3.3  & 26.4  & 45.9  & 33.5  & 83.3  & 75.2  & 233639  & 1217090   \\
Siamese (HDBSCAN) & NA& NA & NA & 3.2  & 9.8   & 4.9  & 1.7  & 19.7  & 3.2  & 26.4  & 44.7  & 33.2  & 84.9  & 73.8  & 237773  & 1279439   \\
Triplet (HDBSCAN) & NA& NA & NA & 3.2  & 10.2  & 4.9  & 1.7  & 20.4  & 3.2  & \textbf{26.4} & \textbf{46.1} & \textbf{33.6} & \textbf{81.0} & 75.6  & 248131  & 1305836 \\ 
 \hline  
\end{tabular}
}

\resizebox{17cm}{!}{%
\begin{tabular}{lcccccccccccccccc}
 \textbf{French}& \multicolumn{3}{c}{\textbf{Grouping}}& \multicolumn{3}{c}{\textbf{Token}}& \multicolumn{3}{c}{\textbf{Type}} & \multicolumn{3}{c}{\textbf{Boundary}}& \multicolumn{4}{c}{\textbf{NLP}} \\ \hline
   & P  & R  & F & P & R & F & P & R & F & P  & R  & F & NED  & Cov & n-words & n-pairs \\ \hline
Baseline  & NA & NA  & NA  & 3.0  & 7.8  & 4.4  & 2.0  & 18.2  & 3.6  & \textbf{28.7} & \textbf{59.4} & \textbf{38.7} & 93.6  & \textbf{89.9} & 199674  & 469019051 \\
Siamese (hybrid)  & 6.7  & 11.1  & 8.4   & 3.2  & 5.3  & 4.0  & \textbf{2.2} & \textbf{15.6} & \textbf{3.8} & 29.2  & 48.0  & 36.3  & 94.5  & 80.7  & 155972  & 252631158 \\
Triplet (hybrid)  & \textbf{8.6} & \textbf{13.8} & \textbf{10.6} & 3.5  & 5.5  & 4.3  & 2.2  & 13.9  & 3.8  & 29.2  & 44.8  & 35.3  & 95.2  & 81.0  & 135774  & 333797926 \\
Siamese (HDBSCAN) & 7.3  & 10.3  & 8.5   & 3.5  & 9.6  & 5.2  & 2.1  & 14.6  & 3.7  & 28.3  & 53.7  & 37.1  & \textbf{88.2} & 83.5  & 148140  & 718587\\
Triplet (HDBSCAN) & 5.9  & 9.9   & 7.4   & \textbf{3.6} & \textbf{9.6} & \textbf{5.2} & 2.2  & 14.6  & 3.8  & 28.3  & 53.6  & 37.1  & 88.6  & 83.3  & 146834  & 710606\\
 \hline  

\end{tabular}
}
\resizebox{17cm}{!}{%
\begin{tabular}{lcccccccccccccccc}
 \textbf{Mandarin}  & \multicolumn{3}{c}{\textbf{Grouping}}& \multicolumn{3}{c}{\textbf{Token}}& \multicolumn{3}{c}{\textbf{Type}} & \multicolumn{3}{c}{\textbf{Boundary}}& \multicolumn{4}{c}{\textbf{NLP}} \\ \hline
    & P  & R  & F & P & R & F & P & R & F & P  & R  & F & NED  & Cov & n-words & n-pairs \\ \hline
Baseline  & 0.4 & 34.6& 0.8 & 2.5 & 2.2 & 2.4 & 2.6 & 4.8 & 3.4 & 31.4& 25.1& 27.9& 93.9 & 65.0  & 15988& 254055  \\
Siamese (hybrid)  & \textbf{11.3}& \textbf{90.8}& \textbf{20.1}& \textbf{3.7} & \textbf{9.1} & \textbf{5.3} & \textbf{4.1} & \textbf{17.1}& \textbf{6.6} & 31.8& 43.1& 36.6& 97.8 & 87.9  & 37396& 358139550  \\
Triplet (hybrid)& 10.1& 85.3& 18.1& 3.5 & 8.6 & 5.0 & 3.8 & 16.3& 6.2 & \textbf{32.0}& \textbf{43.7}& \textbf{36.9}& 97.8 & \textbf{88.9}  & 38044& 238271312  \\
Siamese (HDBSCAN)& 0.5 & 6.4 & 0.9 & 3.5 & 6.5 & 4.6 & 3.7 & 12.9& 5.8 & 31.7& 39.5& 35.2& \textbf{88.7} & 86.8  & 30685& 40773\\
Triplet (HDBSCAN)& 0.4 & 4.6 & 0.7 & 3.4 & 6.9 & 4.6 & 3.6 & 13.6& 5.7 & 31.9& 41.5& 36.1& 90.8 & 88.4  & 33602& 45633  \\ \hline 
\end{tabular}
}
\caption{Performance of the spoken term discovery systems on Zerospeech Challenge 2017 dataset.}
%\caption{Zerospeech Challenge 2017 Track 2 metrics for the spoken term discovery systems on Mandarin dataset.}
\label{table:zerospeech_std}
%\vspace{-0.2cm}
\end{table*}

\subsection{Dataset and evaluation metrics}

Different from the previous work where real life lecture recordings were used but hard to evaluate. Instead, we evaluate the systems with the 2017 Zerospeech Challenge dataset \cite{dunbar2017zero}. The dataset consists of three languages: English, French and Mandarin, with duration of 45 hours, 24 hours and 2.5 hours respectively. 
The challenge provides varies evaluation metrics in the spoken term discovery track. %It includes grouping and type scores that measure the clustering quality, token and boundary scores that measure the parsing quality, and NED and coverage that measure the matching quality of the discovered term clusters.
The grouping, type, token and boundary scores are measured in 3 aspects: precision (\textbf{P}), recall (\textbf{R}) and F-score (\textbf{F}). 
\textbf{Grouping scores} compute the intrinsic quality in terms of the cluster phonetic composition.
\textbf{Type scores} compare how the cluster boundaries match with the actual transcript with true lexicon.
\textbf{Token scores} evaluate the word tokens that are correctly segmented.
\textbf{Boundary scores} evaluate how many actual word boundaries are found.
\textbf{Normalized edit distance (NED)} measures how similar the discovered segments are to the transcript. The smaller the value the more similar. \textbf{Coverage (Cov)} is the fraction of the corpus that are discovered.
There are also metrics that give extra information about the system but are hard to evaluate individually. \textbf{n-words} is the number of system generated spoken term clusters, and \textbf{n-pairs} is the number of segment pairs generated.
Detailed description of each of the metrics can be found here \footnote{Evaluation metrics of the Zerospeech 2017 spoken term discovery track: \url{https://zerospeech.com/2017/track\_2.html}}.

\subsection{Baseline spoken term discovery system}

The same system architecture is adopted as in \cite{sung2018unsupervised}, with slight modification on the system parameters. In the first stage, language dependent ASMs are individually trained on each language. The ASM are trained by first clustering segment-level 40 dimensional multilingual bottleneck features into 55 subwords with Bayesian Gaussian mixtures model (BGMM) \cite{roberts1998bayesian}. The bottleneck features and alignment information are obtained from the same multilingual model trained on 5 corpora. The initial pseudo-transcription of the audio are then trained iteratively with a DNN-HMM to generate the finalized pseudo-transcription. 

In the second stage, word or phrase level segments are obtained from the pseudo-transcription by finding matching subword sequences with local sequence alignment. Normalized Levenshtein distance is used as similarity metric. Different from the previous setup which discards very short segments, all the segments are kept for future use.
The segments are then clustered into spoken term clusters using leader clustering. Ideally the system will run with various system parameters until the optimal language-dependent set is found. But assuming we have no prior knowledge on a zero resource language, a fixed parameters of $T = 0.4, a = 1.8, R= 3$ are used on the three languages. These parameters are determined by prior experience in previous work with slight adjustment. $T$ defines the radius of the cluster, $a$ defines the distance between the center of 2 clusters and $R$ is the minimum subword length of segments being considered for spoken term clustering.

\subsection{Siamese and Triplet network training}
\label{sec:network}
The training data for the Siamese and Triplet networks are created from baseline spoken term clusters as described in Section \ref{sec:conf}. 
The Siamese training pairs and Triplet training tuples are obtained by setting $thres_{\mu_s}$ = 0.2, $thres_{\sigma_s}$= 0.2, and $thres_{\mu_d}$ =  0.4, $thres_{\sigma_d}$ = 0.2. From all the possible valid combinations, $600,000$ Siamese and Triplet training examples are sampled to train the networks respectively.

A simple CNN network is used to construct the sub-network component in the Siamese/Triplet network. The CNN comprises $2$ blocks of Conv-ReLU-Max Pooling followed by a Conv-ReLU connected to $2$ fully-connected layers with a ReLU activation function. The output layer is a linear layer of $40$ dimension.
The input contains frame-level bottleneck features from speech segments. The variable-length feature sequences are zero-padded to derive fixed-length sequences for all segments. %, while the training labels are the similar/dissimilar labels of the input pairs or tuples.
The Siamese and Triplet networks are trained for no more than 20 epochs until a reasonable loss value is attained. After training, the networks are used to transform all segments obtained from the sequence alignment step into fixed-dimension segment representations. Subsequently HDBSCAN is applied to cluster the representations into spoken term clusters. Both HDBSCAN and its hybrid extension are experimented, with cluster selection epsilon being set to $0.2$.
%All the subword sequences obtained during the local sequence alignment stage are then decoded by the Siamese and Triplet networks. 

\section{Results and Analysis}

%\subsection{Zerospeech 2017 challenge result comparison}

Performance of the systems is listed in Table \ref{table:zerospeech_std}. The grouping scores for English are ``NA'', due to the time limit for evaluation is hit and the scoring is not completed. 

First, we look at the coverage of the systems. The baseline system has varying performance on different languages, from 65\% to 89.9\%, showing it is highly language dependent and the optimal model parameters need to be identified individually. However the Siamese and Triplet systems are relatively consistent, with less difference of 75.6\% to 88.4\% on the best system (Triplet HDBSCAN) under the same parameters across all 3 languages. Similar coverage is achieved by Siamese and Triplet networks on both clustering algorithms, with Triplet network being slightly better.
Even the Mandarin baseline system only discovers 65\% of the words, the Siamese and Triplet networks are able to learn effective segment representations that can discover new terms which are not covered before, raising the coverage for at least 20\%. 
However, the coverage of the Siamese and Triplet systems are slightly worse than the baseline by 5\%-10\% on English and French. The missing terms might be the difficult examples that are discovered in the baseline's less confident clusters, but are unable to be well-represented and discovered by the Siamese and Triplet systems. 

As a sanity check, experiments are also done by using the Siamese and Triplet systems to discover only the longer segments (with more than 3 subwords) on English and French, but the coverage is only degraded by at most 3\% on both languages. This implies not much spoken terms can be discovered in shorter segments. One possible reason is that both English and French are more poly-phonetic than Mandarin where the words are likely to be formed by more phonemes. The effect of throwing away short segments is not obvious, beside reducing computation effort.

The number of segment pairs (n-pairs) increases exponentially when data size increases for the baseline. From 25K in Mandarin to 100M-400M in English and French, including too many variations of similar terms. HDBSCAN produces the smallest n-pairs with similar coverage as other systems, which is more preferable. Hybrid generally produces larger n-pairs with similar n-words as HDBSCAN. One possible reason for the small n-pairs in English Triplet hybrid system is that the threshold to activate hybrid mode is not reached yet. 

The proposed systems achieve better cluster quality than the baseline in general, with higher grouping, token, type and boundary scores, especially on Mandarin. HDBSCAN and its hybrid extension have different strength and weakness in terms of grouping quality and NED. HDBSCAN produces lower NED, but gives limited improvement to the grouping scores. While the hybrid method gives a better grouping scores with the exchange of higher NED.

\section{Conclusion}

In this work, the attempt of using Siamese and Triplet networks for spoken term discovery under a complete unsupervised scenario is made. The initial segmentation and cluster information is obtained from other spoken term discovery system. The clusters with high confidence are used to generate matched and mismatched pairs and tuples for training the Siamese and Triplet networks. The networks are used to generate representations for all the available segments, follow by HDBSCAN on the segment representations to obtain new set of spoken term clusters.

It is shown that even the exact labels of the segments are unavailable, Siamese/Triplet network can still be trained when a small set of high confidence matched and mismatched data pairs are presented. %This shows that even in a completely unsupervised scenario, a well-performing Siamese/Triplet network can still be trained with segments and soft labels generated in unsupervised manner. By maintaining a high confidence of hypothesized segment labels, the network is capable to generate segment representations for spoken term discovery. 
Experiments on Zerospeech dataset show that the segment representations generated by Siamese and Triplet networks are effective in representing unseen segments.
Triplet network is slightly better than Siamese network in learning segment representations for spoken term discovery. The proposed systems also provide better cluster quality and consistent coverage than the baseline two-stage model, but still have room for improvement in discovering hard examples.

%Triplet network is less favourable than Siamese network in generating segment representations for spoken term discovery.

%In the problem of spoken term discovery, Triplet network is less favourable in our experiments as the cluster boundaries are less easy to determine by clustering algorithm.

\bibliographystyle{IEEEtran}

\bibliography{main}

% Generated by IEEEtran.bst, version: 1.13 (2008/09/30)
\begin{thebibliography}{10}
\providecommand{\url}[1]{#1}
\csname url@samestyle\endcsname
\providecommand{\newblock}{\relax}
\providecommand{\bibinfo}[2]{#2}
\providecommand{\BIBentrySTDinterwordspacing}{\spaceskip=0pt\relax}
\providecommand{\BIBentryALTinterwordstretchfactor}{4}
\providecommand{\BIBentryALTinterwordspacing}{\spaceskip=\fontdimen2\font plus
\BIBentryALTinterwordstretchfactor\fontdimen3\font minus
  \fontdimen4\font\relax}
\providecommand{\BIBforeignlanguage}[2]{{%
\expandafter\ifx\csname l@#1\endcsname\relax
\typeout{** WARNING: IEEEtran.bst: No hyphenation pattern has been}%
\typeout{** loaded for the language `#1'. Using the pattern for}%
\typeout{** the default language instead.}%
\else
\language=\csname l@#1\endcsname
\fi
#2}}
\providecommand{\BIBdecl}{\relax}
\BIBdecl

\bibitem{wilpon1987investigation}
J.~Wilpon, B.~Juang, and L.~Rabiner, ``An investigation on the use of acoustic
  sub-word units for automatic speech recognition,'' in \emph{ICASSP'87. IEEE
  International Conference on Acoustics, Speech, and Signal Processing},
  vol.~12.\hskip 1em plus 0.5em minus 0.4em\relax IEEE, 1987, pp. 821--824.

\bibitem{lee1988segment}
C.-H. Lee, F.~K. Soong, and B.-H. Juang, ``A segment model based approach to
  speech recognition,'' in \emph{ICASSP-88., International Conference on
  Acoustics, Speech, and Signal Processing}.\hskip 1em plus 0.5em minus
  0.4em\relax IEEE, 1988, pp. 501--541.

\bibitem{park2005towards}
A.~Park and J.~R. Glass, ``Towards unsupervised pattern discovery in speech,''
  in \emph{Automatic Speech Recognition and Understanding, 2005 IEEE Workshop
  on}.\hskip 1em plus 0.5em minus 0.4em\relax IEEE, 2005, pp. 53--58.

\bibitem{versteegh2015zero}
M.~Versteegh, R.~Thiolliere, T.~Schatz, X.~N. Cao, X.~Anguera, A.~Jansen, and
  E.~Dupoux, ``The zero resource speech challenge 2015,'' in \emph{Sixteenth
  annual conference of the international speech communication association},
  2015.

\bibitem{kamper2017embedded}
H.~Kamper, K.~Livescu, and S.~Goldwater, ``An embedded segmental k-means model
  for unsupervised segmentation and clustering of speech,'' in \emph{2017 IEEE
  Automatic Speech Recognition and Understanding Workshop (ASRU)}.\hskip 1em
  plus 0.5em minus 0.4em\relax IEEE, 2017, pp. 719--726.

\bibitem{thual2018k}
A.~Thual, C.~Dancette, J.~Karadayi, J.~Benjumea, and E.~Dupoux, ``A k-nearest
  neighbours approach to unsupervised spoken term discovery,'' in \emph{2018
  IEEE Spoken Language Technology Workshop (SLT)}.\hskip 1em plus 0.5em minus
  0.4em\relax IEEE, 2018, pp. 491--497.

\bibitem{jansen2011efficient}
A.~Jansen and B.~Van~Durme, ``Efficient spoken term discovery using randomized
  algorithms,'' in \emph{2011 IEEE Workshop on Automatic Speech Recognition \&
  Understanding}.\hskip 1em plus 0.5em minus 0.4em\relax IEEE, 2011, pp.
  401--406.

\bibitem{harwath2013zero}
D.~F. Harwath, T.~J. Hazen, and J.~R. Glass, ``Zero resource spoken audio
  corpus analysis,'' in \emph{2013 IEEE International Conference on Acoustics,
  Speech and Signal Processing}.\hskip 1em plus 0.5em minus 0.4em\relax IEEE,
  2013, pp. 8555--8559.

\bibitem{siu2014unsupervised}
M.-h. Siu, H.~Gish, A.~Chan, W.~Belfield, and S.~Lowe, ``Unsupervised training
  of an hmm-based self-organizing unit recognizer with applications to topic
  classification and keyword discovery,'' \emph{Computer Speech \& Language},
  vol.~28, no.~1, pp. 210--223, 2014.

\bibitem{sung2018unsupervised}
M.-L. Sung, S.~Feng, and T.~Lee, ``Unsupervised pattern discovery from thematic
  speech archives based on multilingual bottleneck features,'' in \emph{2018
  Asia-Pacific Signal and Information Processing Association Annual Summit and
  Conference (APSIPA ASC)}.\hskip 1em plus 0.5em minus 0.4em\relax IEEE, 2018,
  pp. 1448--1455.

\bibitem{bromley1994signature}
J.~Bromley, I.~Guyon, Y.~LeCun, E.~S{\"a}ckinger, and R.~Shah, ``Signature
  verification using a" siamese" time delay neural network,'' in \emph{Advances
  in neural information processing systems}, 1994, pp. 737--744.

\bibitem{koch2015siamese}
G.~Koch, R.~Zemel, and R.~Salakhutdinov, ``Siamese neural networks for one-shot
  image recognition,'' in \emph{ICML deep learning workshop}, vol.~2, 2015.

\bibitem{hoffer2015deep}
E.~Hoffer and N.~Ailon, ``Deep metric learning using triplet network,'' in
  \emph{International Workshop on Similarity-Based Pattern Recognition}.\hskip
  1em plus 0.5em minus 0.4em\relax Springer, 2015, pp. 84--92.

\bibitem{kamper2016deep}
H.~Kamper, W.~Wang, and K.~Livescu, ``Deep convolutional acoustic word
  embeddings using word-pair side information,'' in \emph{2016 IEEE
  International Conference on Acoustics, Speech and Signal Processing
  (ICASSP)}.\hskip 1em plus 0.5em minus 0.4em\relax IEEE, 2016, pp. 4950--4954.

\bibitem{svec2017relevance}
J.~V. Psutka and J.~Trmal, ``A relevance score estimation for spoken term
  detection based on rnn-generated pronunciation embeddings.'' in
  \emph{INTERSPEECH}, 2017, pp. 2934--2938.

\bibitem{zhu2018siamese}
Z.~Zhu, Z.~Wu, R.~Li, H.~Meng, and L.~Cai, ``Siamese recurrent auto-encoder
  representation for query-by-example spoken term detection.''

\bibitem{bhati2019unsupervised}
S.~Bhati, S.~Nayak, K.~S.~R. Murty, and N.~Dehak, ``Unsupervised acoustic
  segmentation and clustering using siamese network embeddings.'' 2019.

\bibitem{riad2018sampling}
R.~Riad, C.~Dancette, J.~Karadayi, N.~Zeghidour, T.~Schatz, and E.~Dupoux,
  ``Sampling strategies in siamese networks for unsupervised speech
  representation learning,'' \emph{arXiv preprint arXiv:1804.11297}, 2018.

\bibitem{campello2013density}
R.~J. Campello, D.~Moulavi, and J.~Sander, ``Density-based clustering based on
  hierarchical density estimates,'' in \emph{Pacific-Asia conference on
  knowledge discovery and data mining}.\hskip 1em plus 0.5em minus 0.4em\relax
  Springer, 2013, pp. 160--172.

\bibitem{malzer2019hybrid}
C.~Malzer and M.~Baum, ``A hybrid approach to hierarchical density-based
  cluster selection,'' \emph{arXiv preprint arXiv:1911.02282}, 2019.

\bibitem{dunbar2017zero}
E.~Dunbar, X.~N. Cao, J.~Benjumea, J.~Karadayi, M.~Bernard, L.~Besacier,
  X.~Anguera, and E.~Dupoux, ``The zero resource speech challenge 2017,'' in
  \emph{2017 IEEE Automatic Speech Recognition and Understanding Workshop
  (ASRU)}.\hskip 1em plus 0.5em minus 0.4em\relax IEEE, 2017, pp. 323--330.

\bibitem{roberts1998bayesian}
S.~J. Roberts, D.~Husmeier, I.~Rezek, and W.~Penny, ``Bayesian approaches to
  gaussian mixture modeling,'' \emph{IEEE Transactions on Pattern Analysis and
  Machine Intelligence}, vol.~20, no.~11, pp. 1133--1142, 1998.

\end{thebibliography}

\end{document}